\documentclass[letter]{aa}
\usepackage{txfonts,epsfig}

\begin{document}
\title{Zooming in on Supernova 1987A at sub-mm wavelengths}
\author{Ma\v{s}a Laki\'cevi\'c\inst{1,2}
\and    Jacco Th.\ van Loon\inst{2}
\and    Thomas Stanke\inst{1}
\and    Carlos De Breuck\inst{1}
\and    Ferdinando Patat\inst{1}
}
\institute{European Organization for Astronomical Research in the Southern
           Hemisphere (ESO), Karl-Schwarzschild-Str.\ 2, D-85748, Garching b.\
           M\"unchen, Germany, \email{mlakicev@eso.org}
\and       Astrophysics Group, Lennard-Jones Laboratories, Keele University,
           Staffordshire ST5 5BG, UK}
\date{
Submitted: 16 December 2011;
Final version: 14 March 2012
}
\abstract
{Supernova 1987A (SN\,1987A) in the neighbouring Large Magellanic Cloud offers
a superb opportunity to follow the evolution of a supernova and its remnant in
unprecedented detail. Recently, far-infrared (far-IR) and sub-mm emission was
detected from the direction of SN\,1987A, which was interpreted as due to the
emission from dust, possibly freshly synthesized in the SN ejecta.}
{To better constrain the location and hence origin of the far-IR and
sub-mm emission in SN\,1987A, we have attempted to resolve the object in that
part of the electro-magnetic spectrum.}
{We observed SN\,1987A during July--September 2011 with the Atacama Pathfinder
EXperiment (APEX), at a wavelength of 350 $\mu$m with the Submillimetre APEX
Bolometer CAmera (SABOCA) and at 870 $\mu$m with the Large APEX BOlometer
CAmera (LABOCA). The 350-$\mu$m image has superior angular resolution
($8^{\prime\prime}$) over that of the Herschel Space Observatory 350-$\mu$m
image ($25^{\prime\prime}$). The 870-$\mu$m observation (at
$20^{\prime\prime}$ resolution) is a repetition of a similar observation made in
2007.}
{In both images, at 350 and 870 $\mu$m, emission is detected from SN\,1987A,
and the source is unresolved. The flux densities in the
new (2011) measurements are consistent with those measured before with
Herschel at 350 $\mu$m (in 2010) and with APEX at 870 $\mu$m (in 2007). A
higher dust temperature ($\approx33$ K) and lower dust mass might be possible
than what was previously thought.}
{The new measurements, at the highest angular resolution achieved so far at
far-IR and sub-mm wavelengths, strengthen the constraints on the location of
the emission, which is thought to be close to the site of SN\,1987A and its circumstellar ring
structures. These measurements set the stage for upcoming observations at even
higher angular resolution with the Atacama Large Millimeter Array (ALMA).}
\authorrunning{Laki\'cevi\'c et al.}
\titlerunning{Zooming in on Supernova 1987A at sub-mm wavelengths}
\maketitle
\section{Introduction}

SN\,1987A offers us the unique opportunity to follow the early evolution of a
supernova (SN) and its transition into a supernova remnant (SNR) in exquisite
detail. Located in the Large Magellanic Cloud (LMC) at a relatively accurately
known distance of 50 kpc, it is seen in a direction that suffers from little
interstellar extinction. While SN\,1987A was classified as type II (Arnett et
al.\ 1989), it showed distinct properties with respect to typical plateau
events, most probably linked to its unexpected progenitor star: the blue
supergiant Sk\,$-69^\circ$\,202.

SN\,1987A's emerging remnant is being shaped by the impact of the initial
blastwave and the expanding hot plasma upon an equatorial ring of matter,
$1\rlap{.}^{\prime\prime}6$ in diameter, deposited probably by the supernova
progenitor star when it was still a red supergiant (Bouchet al.\ 2006; Dwek et
al.\ 2010; Larsson et al.\ 2011). The ejecta themselves are due to reach the
ring in the near future. The increasing strength of the shock is also
reflected in an exponential increase in radio brightness (Zanardo et al.\
2010).

Small amounts ($M_{\rm d}\leq10^{-3}$ M$_\odot$) of warm dust formed in about
two years since the explosion (Suntzeff \& Bouchet 1990; Bouchet, Danziger \&
Lucy 1991). This offers a great opportunity to study fresh supernova dust, because
in older remnants it becomes increasingly difficult to distinguish between
dust formed in the ejecta or dust swept-up from the surroundings (e.g., Rho et
al.\ 2008, 2009; Sandstrom et al.\ 2009; van Loon et al.\ 2010; Otsuka et al.\
2010; Barlow et al.\ 2010). Early detection also allows one to monitor the
evolution of the dust, which might be altered or even destroyed before
entering the interstellar medium. Claims of observaions of dust formation have been made in
more distant SNe (e.g., Elmhamdi et al.\ 2003; Fox et al.\ 2009; Sakon et al.\
2009; Inserra et al.\ 2011), though often SNe are seen to destroy pre-existing
dust (e.g., Botticella et al.\ 2009; Wesson et al.\ 2010; Andrews et al.\
2011). SN\,1987A offers prospects of locating the cold dust directly, by
spatially resolving its emission.

Far-IR emission was detected from SN\,1987A with the {\it Herschel} Space
Observatory by Matsuura et al.\ (2011), who suggested a colder dust component
(17--23 K) than that in the equatorial ring ($\approx170$ K). They placed
the cold dust within the ejecta, to reconcile the large mass of 0.5 M$_\odot$ they derive. 
This would imply enormous growth since the
earlier detections. However, the {\it Herschel} images do not resolve the
equatorial ring from the ejecta, and still allow for the dust to reside
outside the ring -- be it of progenitor origin or interstellar dust echoing
the supernova (Meikle et al.\ 2011). We also remind the reader that chance
superpositions do occur (e.g., van Loon \& Oliveira 2003).

SN\,1987A was observed at sub-mm wavelengths (870 $\mu$m) with the Large APEX
BOlometer CAmera (LABOCA; Siringo et al.\ 2009) on the Atacama Pathfinder
EXperiment (APEX) for the first time in 2007 (Laki\'cevi\'c et al.\ 2011). We
observed this object again in 2011 with LABOCA at 870 $\mu$m, and at 350
$\mu$m with SABOCA (Submillimetre APEX Bolometer CAmera; Siringo et al.\
2010), at three times higher angular resolution than the {\it Herschel}
observations at that wavelength. We here present the results. These new
observations place important constraints on the maximum spatial extent of the
dust emission, and they provide an important zero-spacing value for planned
observations with the Atacama Large Millimeter Array (ALMA).

\section{Observations}

Observations of SN\,1987A (programme ESO\,088.D-0252, P.I.\ Ma\v{s}a
Laki\'cevi\'c) were made with APEX, a 12-m single-dish sub-mm telescope at the
ALMA site at an altitude of 5100 m. We used the bolometer array cameras LABOCA
and SABOCA, in July and September 2011 under excellent conditions -- the water
vapour column was $<0.2$ mm for most of the SABOCA observations and $<0.5$ mm
for the LABOCA observations.

\subsection{LABOCA observations at 870 $\mu$m}

SN\,1987A was observed with LABOCA at a wavelength of 870 $\mu$m (345 GHz)
within a 60-GHz wide band ($\Delta\nu/\nu\approx1/6$). The on-source
integration time was 5 h (6.1 h including overheads). The field was imaged in
spiral raster map mode. The beam has a full-width at half-maximum (FWHM) of
$\sim19.2^{\prime\prime}$.

The data were reduced with the bolometer array analysis ({\sc BoA}) software.
The data were first corrected for atmospheric attenuation determined through
``skydips'', applying a correction derived from observations of planets and
secondary flux calibrators taken at similar elevation as the target. Then
several rounds of correlated flux variation (``skynoise'') removal, despiking,
removal of noisy bolometers, and baseline subtractions were applied, before
maps of the individual scans were created. These were then combined. The
resulting map was then smoothed, and values below zero set to zero, and used
as a source model for the next iteration. The model was subtracted from the
data before skynoise, spike and baseline removal and added back before
creating the next map. In total 35 iterations were performed, with increasing
skynoise, spike and baseline removal settings while reducing the smoothing.
This procedure allows one to recover extended structures, reduces the negative
``bowls'' seen frequently in bolometer on-the-fly maps, and reduces the noise,
while yielding a final map at the full spatial resolution of the instrument.

%
%
\begin{figure}
\centerline{\psfig{figure=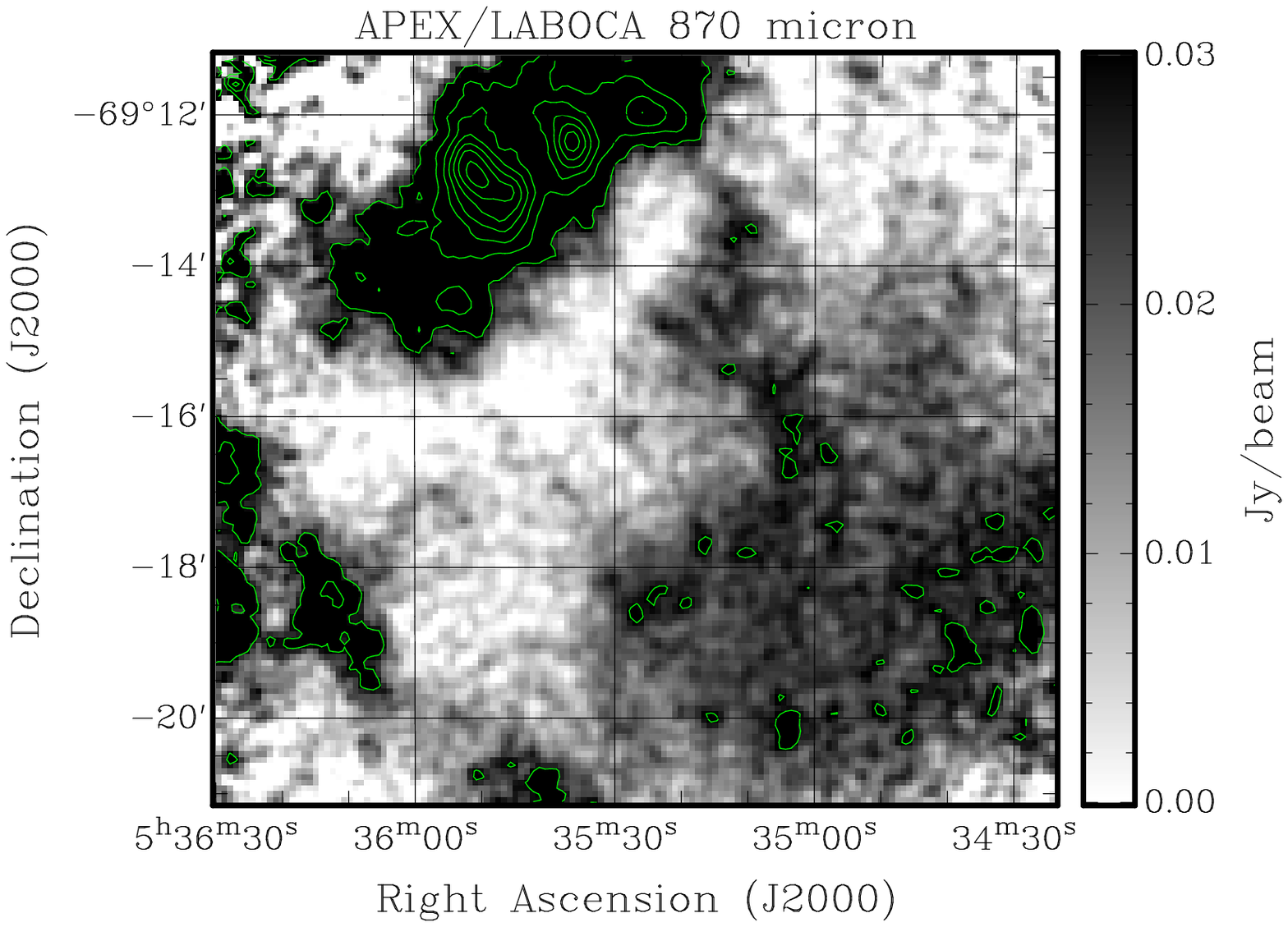,width=90mm}}
\caption[]{APEX/LABOCA image of SN\,1987A at 870 $\mu$m, beam FWHM
$20\rlap{.}^{\prime\prime}4$. Contours are used where the brightness exceeds
30 mJy beam$^{-1}$. SN\,1987A is located at $5^{\rm h}35^{\rm
m}28\rlap{.}^{\rm s}0$, $-69^\circ16^\prime11^{\prime\prime}$.}
\label{LABOCABoA}
\end{figure}

The image was smoothed with half a beam; the resulting beam size was
$20\rlap{.}^{\prime\prime}4$. It is presented in full in Fig.\ 1, and is zoomed-in on the position of SN\,1987A in Fig.\ 2 (top right). The source appears to
sit on filamentary emission extending towards the south--west, similar to that
reported by Laki\'cevi\'c et al.\ (2011).

The flux density of the central point source is $F_{870}=19.6\pm5.6$ mJy,
where the uncertainty combines the r.m.s.\ noise level of 4 mJy (8 mJy in the
unsmoothed map) and LABOCA's absolute calibration uncertainty of $\sim20$\%.
This brightness is consistent within the errors with the result from
Laki\'cevi\'c et al.\ (2011), viz.\ $21\pm4$ mJy, which was obtained from
measurements in 2007.

%
\begin{figure*}
\centerline{
\vbox{
\hbox{
\hspace{1mm}
\psfig{figure=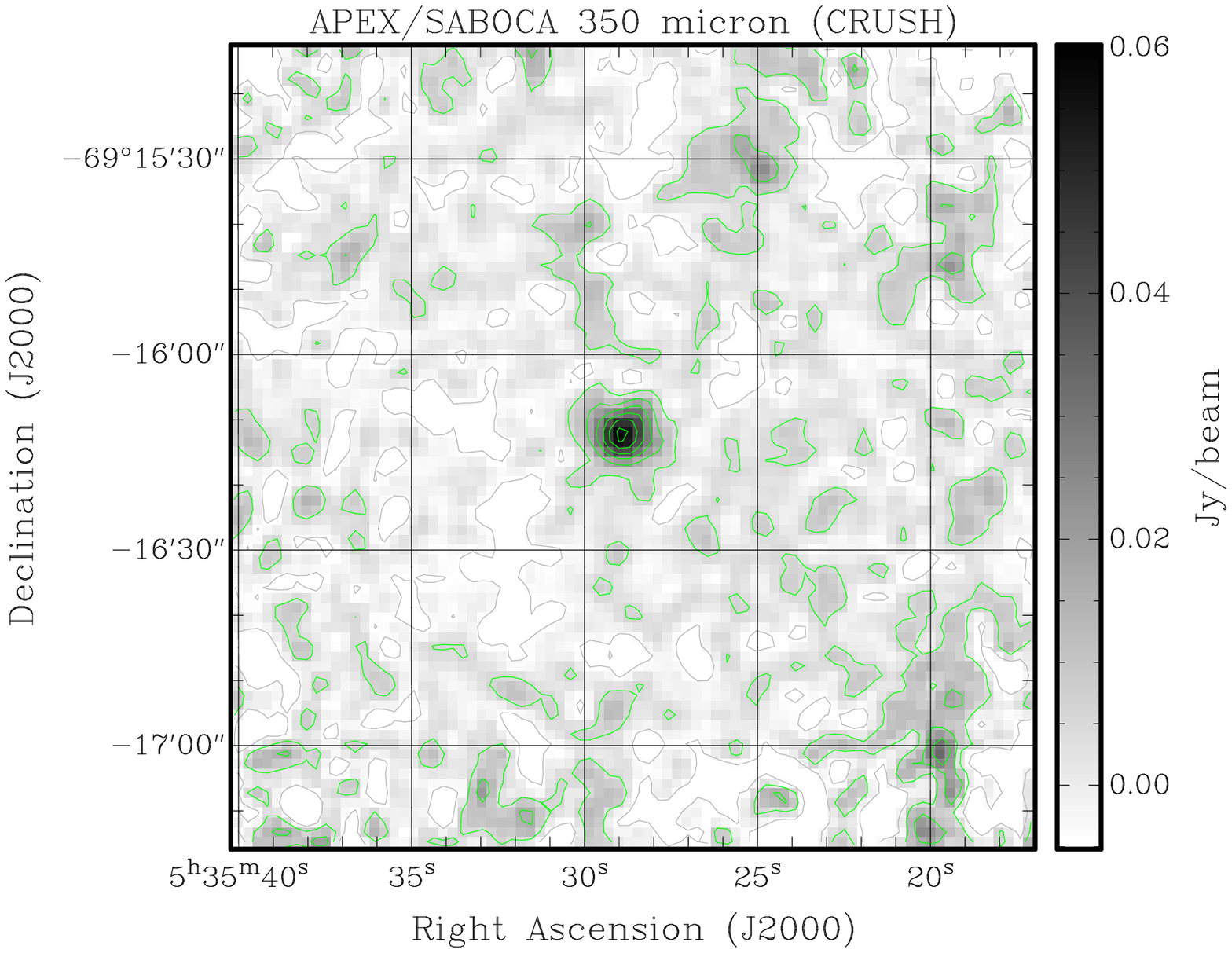,width=60mm}
\hspace{1mm}
\psfig{figure=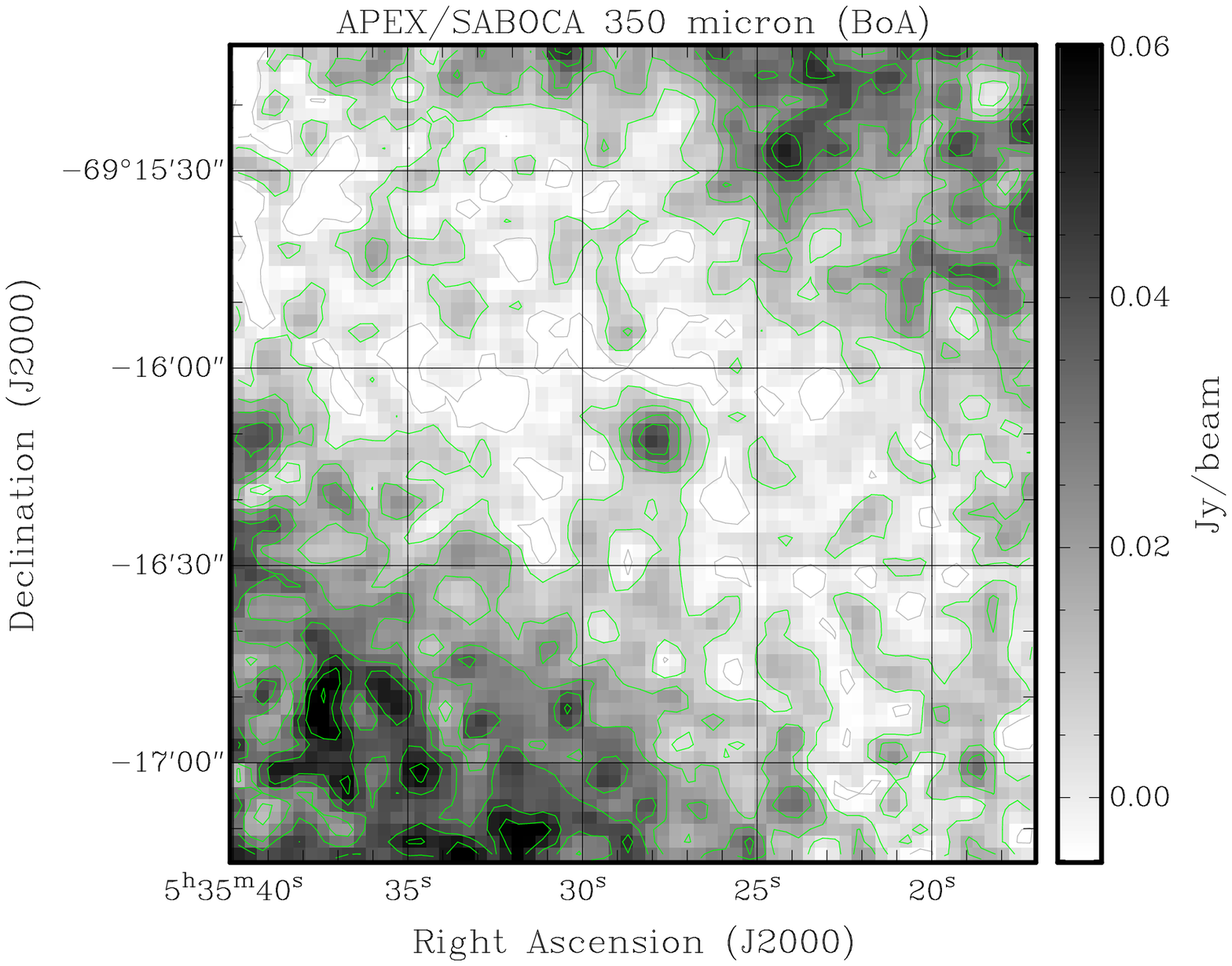,width=59mm}
\hspace{1mm}
\psfig{figure=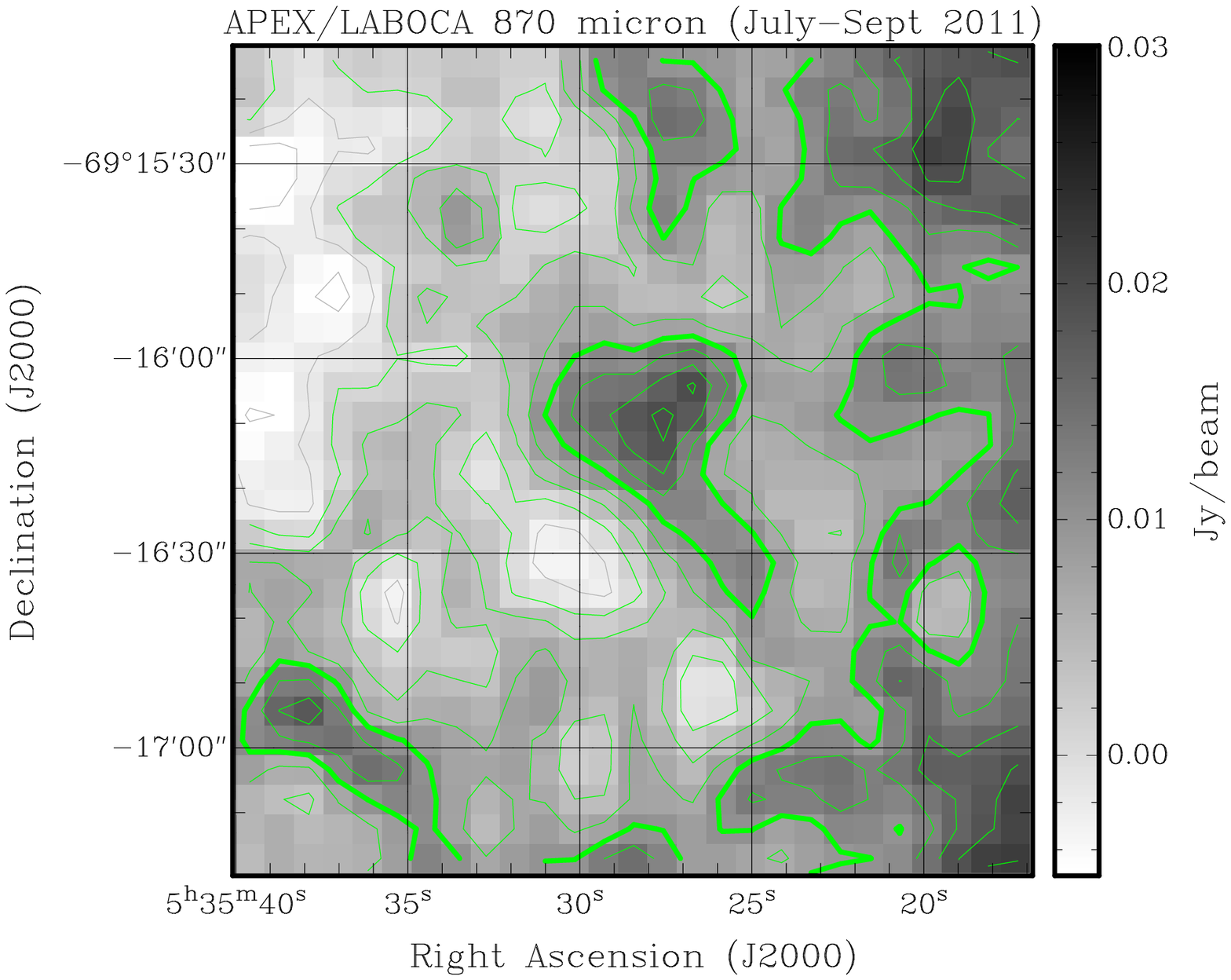,width=58mm}
}
\vspace{1mm}
\hbox{
\hspace{0mm}
\psfig{figure=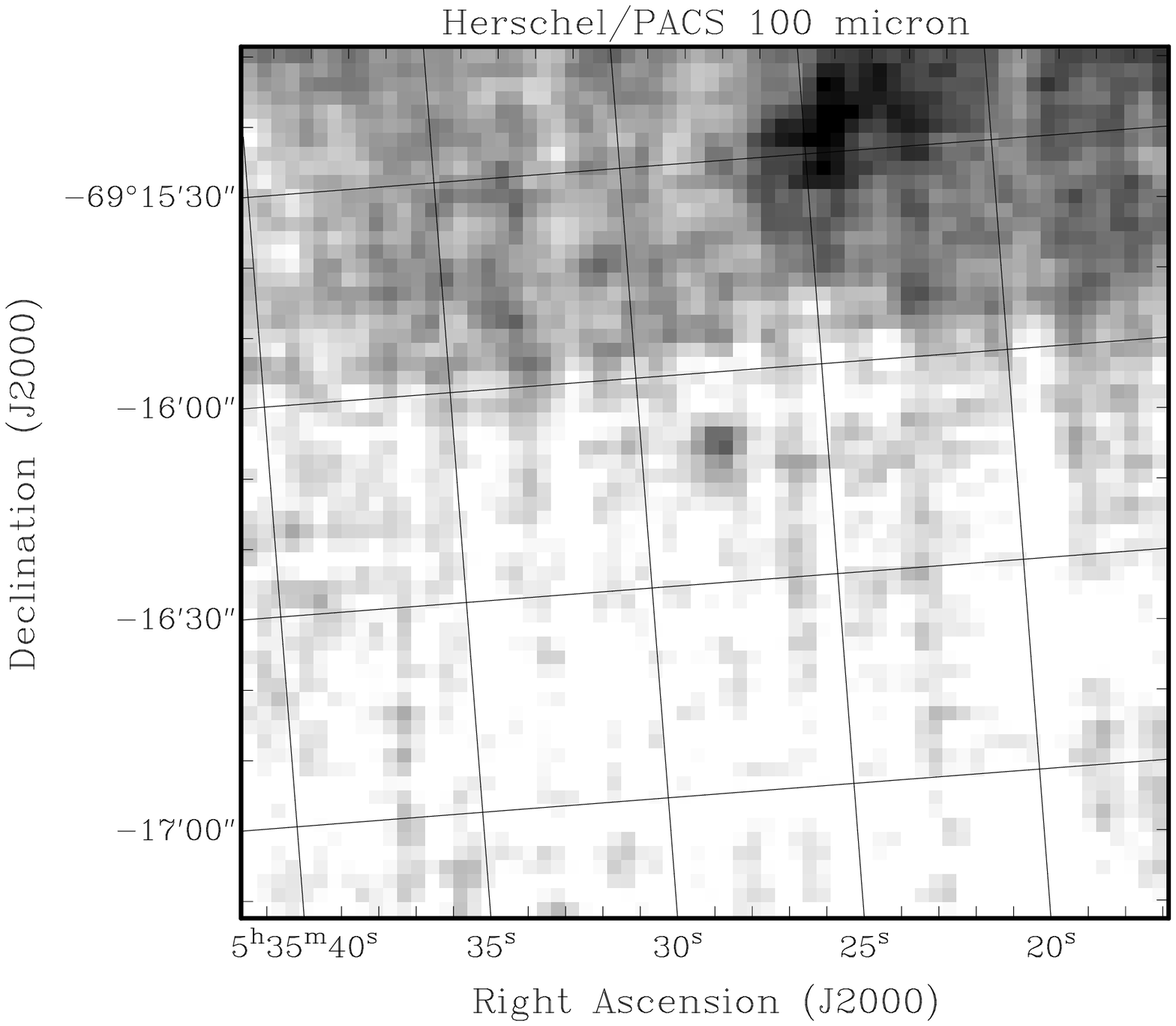,width=53mm}
\hspace{6mm}
\psfig{figure=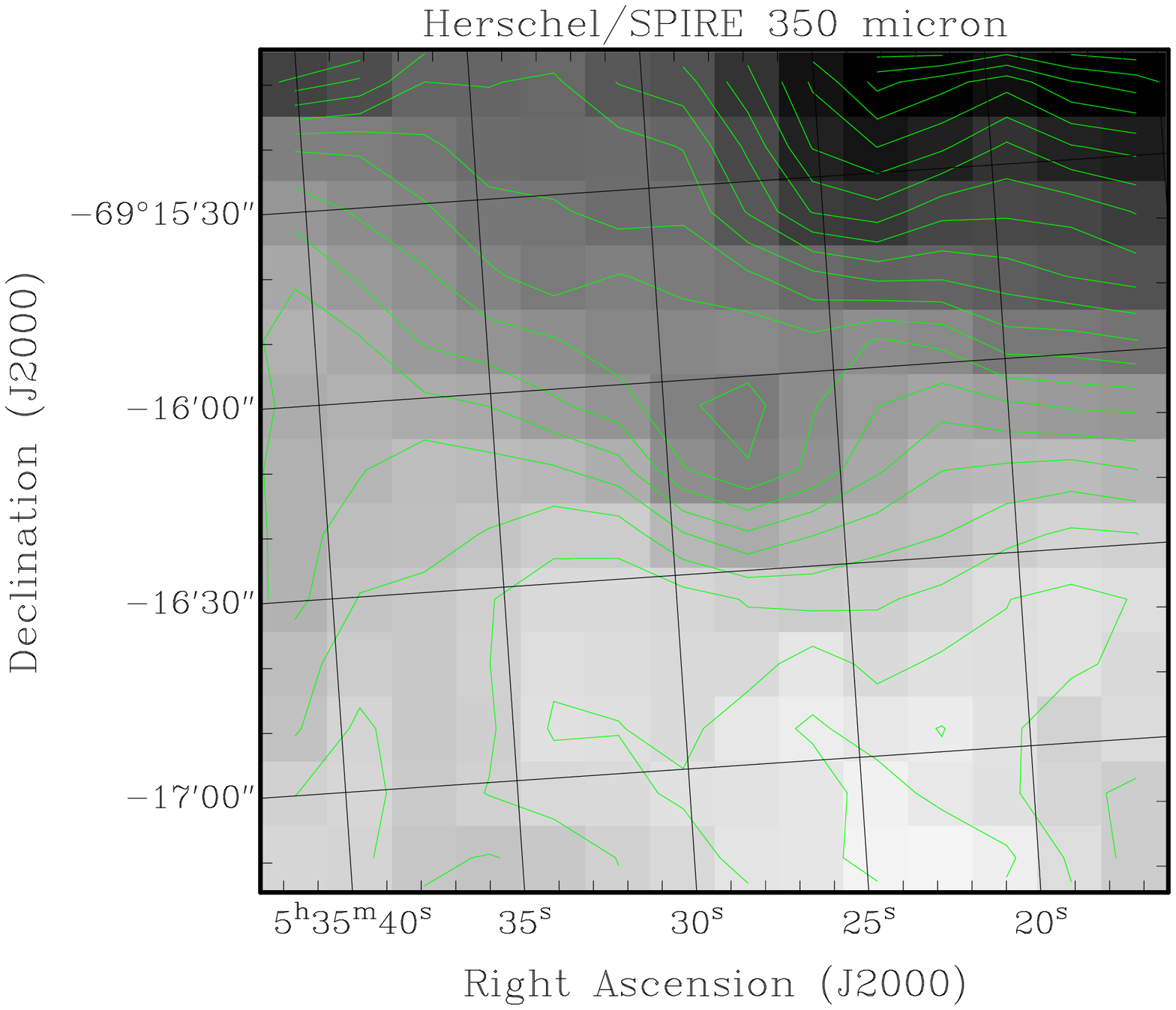,width=54.5mm}
\hspace{8mm}
\psfig{figure=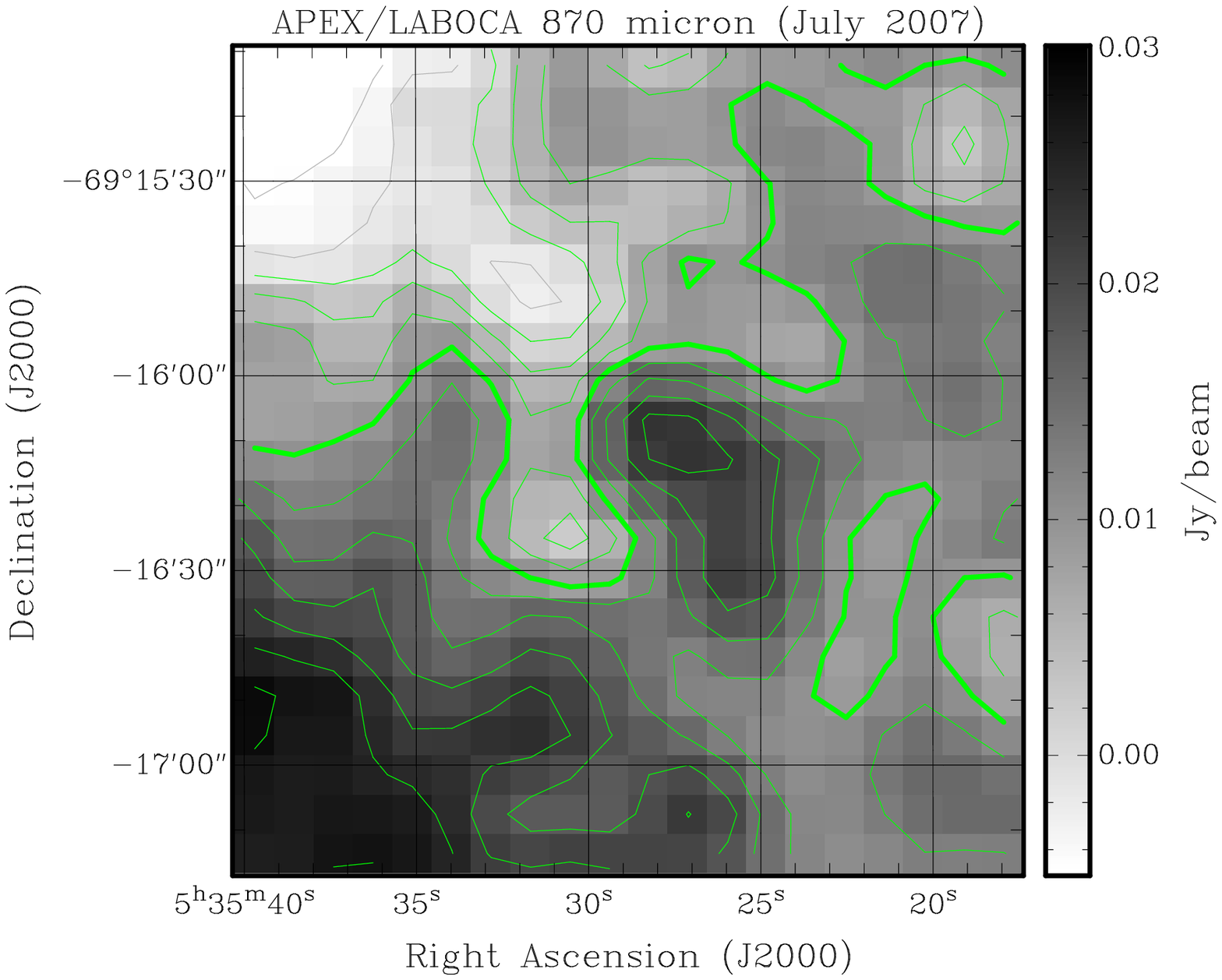,width=58mm}
}
}
}
\caption[]{{\it Top:} APEX images of SN\,1987A -- {\it left:} at 350 $\mu$m
reduced with {\sc crush}, FWHM $8\rlap{.}^{\prime\prime}4$; {\it centre:} at
350 $\mu$m reduced with {\sc BoA}, FWHM $8\rlap{.}^{\prime\prime}4$; {\it
right:} at 870 $\mu$m, FWHM $20\rlap{.}^{\prime\prime}4$. {\it Bottom:}
previous images from {\it Herschel} (Matsuura et al.\ 2011) and APEX
(Laki\'cevi\'c et al.\ 2011) -- {\it left:} at 100 $\mu$m, FWHM
$6\rlap{.}^{\prime\prime}8$; {\it centre:} at 350 $\mu$m, FWHM
$24\rlap{.}^{\prime\prime}9$; {\it right:} at 870 $\mu$m with APEX, FWHM
$23^{\prime\prime}$. SN\,1987A is located at $5^{\rm h}35^{\rm
m}28\rlap{.}^{\rm s}0$, $-69^\circ16^\prime11^{\prime\prime}$.}
\label{SABOCALABOCA}
\end{figure*}

\subsection{SABOCA observations at 350 $\mu$m}

SABOCA is a continuum receiver that consists of 39 bolometers operating at a wavelength of 350 $\mu$m (857 GHz)
within a 120-GHz wide band ($\Delta\nu/\nu\approx1/7$). The on-source
integration time was 15.6 h (22.7 h including overheads). The field was imaged
in spiral raster map mode. The beam has a FWHM of
$7\rlap{.}^{\prime\prime}8\pm0\rlap{.}^{\prime\prime}5$.

The data reduction was performed with two different packages -- {\sc BoA} and
{\sc crush} Comprehensive Reduction Utility for SHARC-2; Kov\'acs 2008 --
which use different algorithms.

In {\sc crush}, we used the faint source reduction option, with a source size
filtering of $10^{\prime\prime}$, and ran 20 iterations. The image, smoothed
with half a beam, is presented in Fig.\ 2 (top left). Since the pointing in
September was slightly offset from that in July, the data from July and September
were reduced separately and then the September data were shifted to match the
July data and averaged. The resulting beam has an effective FWHM of
$8\rlap{.}^{\prime\prime}4$. The flux density of the unresolved source at the
position of SN\,1987A on the combined image is $F_{350,\ {\rm CRUSH}}=58\pm13$
mJy, where the uncertainty combines the r.m.s.\ noise level of 6 mJy and
SABOCA's absolute calibration uncertainty of $\sim20$\%.

The data reduction strategy within {\sc BoA} corresponds to the one used for
LABOCA, except that the atmospheric opacities were determined from
measurements of the amount of precipitable water vapour (PWV) done in parallel
to the observations, using an atmospheric model (ATM). Because the calibration
measurements were made at the same time as the observations and at the same
azimuth and elevation, they yield a more accurate opacity correction. The
image obtained using {\sc BoA}, smoothed with half a beam, is presented in
Fig.\ 2 (top, centre). Here, the July and September data were aligned and
placed at the centre of the image {\it before} the reduction. The measured
flux density of SN\,1987A is now $F_{350,\ {\rm BoA}}=44\pm7$ mJy.

The maps reduced with {\sc BoA} and {\sc crush} differ slightly in the flux
level, probably due to the different calibration strategies ({\sc crush} uses
opacities derived from skydips). The flux densities are consistent within the
calibration uncertainties of $\sim20$--30\%. The extended emission in the {\sc
BoA} reduced image is not evident in the {\sc crush} reduced map, because here the
algorithm is optimised to remove correlated signal variations from the data
time streams, which includes extended emission structures.

The images from the two reduction techniques both confirm that the object at
the position of SN\,1987A is a point source. The detection is of much higher
significance than the uncertainty in the flux density suggests, because the latter
includes systematic uncertainties mainly in the flux calibration. For the flux
density we adopt the average from the two reductions, i.e.\ $F_{350}=51\pm10$
mJy, where the uncertainty is the average from the individual uncertainties as
they are strongly correlated. This is remarkably close to the {\it Herschel}
measurement, viz.\ $49.3\pm6.5$ mJy, in 2010.

\section{Discussion}

Models predict $\sim0.1$--1 M$_\odot$ of dust to form in the ejecta of
core-collapse SNe (Nozawa et al. 2003). Warm dust ($\approx170$ K) was
identified with the equatorial ring around SN\,1987A by direct imaging
(Bouchet et al.\ 2006). Dust formed within the ejecta might be colder. The
cooling rate of a spherical grain of diameter $a$ and temperature $T_{\rm d}$
is its luminosity, $L_{\rm d}=4\pi(a/2)^2\sigma T_{\rm d}^4$, resulting in a
heat loss $-m C dT_{\rm d}/dt$ where $m$ is the mass of the grain and $C$ its
specific heat. Integrating, noting that the initial temperature rapidly
vanishes, leads to
\begin{equation}
T_{\rm d}\simeq \left(\frac{a\ \rho\ C}{18\ \sigma}\right)^{1/3} t^{-1/3},
\end{equation}
where $\rho$ is the grain's density. For $\rho\sim3$ g cm$^{-3}$, $C\sim1$ J
K$^{-1}$ g$^{-1}$ and $a=0.1$ $\mu$m one finds that the grain cools to the
temperature of the cosmic microwave background (3 K) within a matter of hours.
First, collisions with the gas prevent this from happening, but as the ejecta
expand the gas cools and the gas and grains detach (when the density falls
below $\sim10^5$ cm$^{-3}$). The interstellar radiation field (ISRF) keeps the
dust at a temperature $10<T_{\rm d}<30$ K. This led Matsuura et al.\ (2011) to
suggest that the $\sim20$-K dust attributed to the far-IR emission from
SN\,1987A formed within the SN ejecta.


\subsection{Spectral energy distribution (SED)}

In Fig.\ 3 we present the SED of SN\,1987A. Laki\'cevi\'c et al.\ (2011)
described the far-IR/sub-mm emission using a modified black body with a dust
emission coefficient $\kappa/\kappa_0=(\lambda/\lambda_0)^\alpha$; they found
$\alpha=-1.5$ and $T_{\rm d}\approx18$ K, similar to the values reported in
Matsuura et al.\ (2011) and hence a similar dust mass ($\approx0.5$ M$_\odot$)
depending on $\kappa_0$. Synchrotron emission contributes $\sim5$ and $\sim9$
mJy at 0.35 and 0.87 mm, respectively, as estimated from extrapolation of a
powerlaw $F_\nu\propto\lambda^\alpha$ with slope $\alpha=+0.7$ (Fig.\ 3).
Limits on the contribution from free--free emission at those wavelengths are
rather more stringent (Laki\'cevi\'c et al.\ 2011).

%
\begin{figure}
\centerline{\psfig{figure=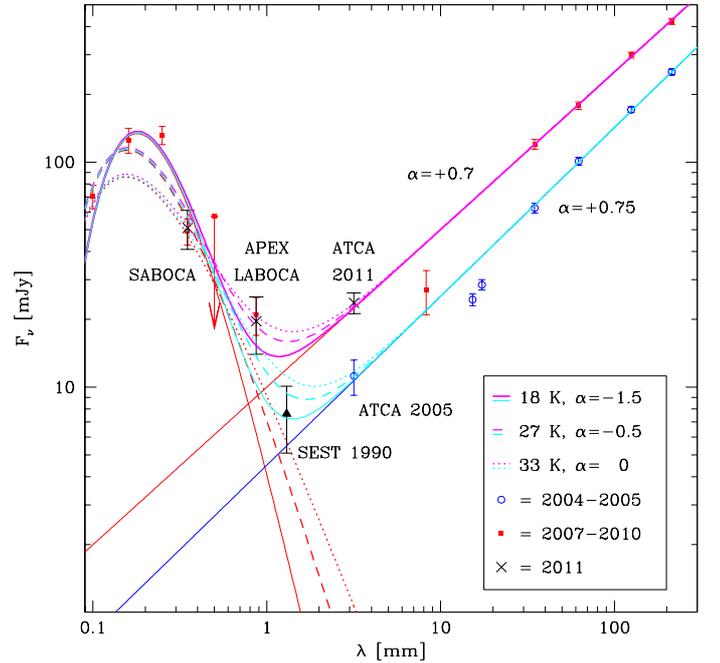,width=90mm}}
\caption[]{SED of SN\,1987A. Data: Biermann et al.\ (1992); Zanardo et al.\
(2010); Matsuura et al.\ (2011); Laki\'cevi\'c et al.\ (2011, 2012); this
work.}
\label{SED}
\end{figure}

The new measurements at 0.35 and 0.87 mm show no change from the previous
measurements at those wavelengths with {\it Herschel} (Matsuura et al.\ 2011;
data from August 2010) and APEX (Laki\'cevi\'c et al.\ 2011; data from July
2007), respectively. Hence there is no evidence for an evolution of the dust on
those timescales, be it the amount, temperature, or grain sizes.

However, values of $\alpha$ closer to zero may not be unexpected at sub-mm
wavelengths (Paradis et al.\ 2011). Adopting $\alpha=0$, and raising the dust
temperature to 33 K results in a better fit to the 0.1, 0.35- and 0.87-mm
measurements (Fig.\ 3), leaving only the {\it Herschel} 160- and 250-$\mu$m
measurements discrepant. A compromise is obtained for $\alpha=-0.5$, at 27 K,
which is at the extreme end of the models in Fig.\ 7 of Paradis et al.\
(2011); this reconciles the 160-$\mu$m datum (Fig.\ 3). The {\it Herschel}
data could be affected by line emission close to SN\,1987A: the 160-$\mu$m
band includes the [C\,{\sc ii}] 158-$\mu$m and [N\,{\sc ii}] 205-$\mu$m lines,
while the 250-$\mu$m band includes three highly excited $^{12}$CO lines with
another $^{12}$CO line and the [N\,{\sc ii}] line at the edges of the band. In
warm gas, these $^{12}$CO transitions can be brighter than those at longer
wavelengths, and the 350- and 500-$\mu$m bands each only include two $^{12}$CO
lines, which additionally reduces the contribution from line emission at $\lambda>300$
$\mu$m. The fit for $\alpha=0$ (and, though to a lesser degree, the fit for
$\alpha=-0.5$) agrees with the measurement at 1.3 mm made in 1990 (Biermann et
al.\ 1992), hinting that the cold dust has been unaltered since day 1290 and
that it might well have already been in place prior to the explosion.
Following Evans et al.\ (2003) we obtain
\begin{equation}
M_{\rm d}/{\rm M}_\odot = 1.5\times10^3\kappa_{1{\rm mm}}^{-1},
\end{equation}
where $\kappa_\lambda$ cannot exceed the ratio of grain surface area and grain
mass. Hence, a spherical black body with $a=0.1$ $\mu$m and $\rho=3$ g
cm$^{-3}$ would yield $\kappa=2\times10^5$ cm$^2$ g$^{-1}$. This would reduce the mass estimate to $M_{\rm d}\approx0.008$ M$_\odot$ and within the
realms of the formation through mass loss from the SN progenitor (assuming
several M$_\odot$ were lost and a gas-to-dust ratio of a few hundred). For
$\alpha=-0.5$ or $-1.5$ at $\lambda>100$ $\mu$m we would have found $M_{\rm
d}\approx0.024$ or 0.24 M$_\odot$, respectively. Mennella et al.\ (1998) find
$\kappa\sim10$--$10^3$ cm$^2$ g$^{-1}$ at far-IR--mm wavelengths, depending on
grain composition, but smaller grains, needles, or fluffy grains all increase
the surface-to-mass ratio and hence $\kappa$. Indeed, for SNe of type IIb --
probably a good match to SN\,1987A -- Nozawa et al.\ (2010) predict grains of
only 0.001--0.003 $\mu$m size, and Wickramasinghe \& Wickramasinghe (1993)
explained the mm radiation from SN\,1987A in 1990 by iron whiskers with
$\kappa\sim10^7$ cm$^2$ g$^{-1}$ at far-IR--mm wavelengths.

\subsection{Where is the dust?}

Our APEX 350-$\mu$m image of SN\,1987A has a beam size (area) a tenth of that
of the {\it Herschel} 350-$\mu$m image (Matsuura et al.\ 2011). The emission
from SN\,1987A remains unresolved, i.e.\ it is located {\it well} within
$8^{\prime\prime}$ (corresponding to 2 pc at the distance of the LMC of 50
kpc) from the explosion site. Even if it were due to a light echo (Meikle et
al.\ 2011) this suggests the echoing dust is probably situated $<2$ pc from
the explosion site and was most likely produced by the progenitor of
SN\,1987A. Little room now remains for cold dust to surround the equatorial
ring, i.e.\ in the dense wind material from the progenitor; the dust is more
likely to reside within the ring or inside of it (e.g., in the ejecta).

The grains might be heated to 33 K by the decay of $^{44}$Ti within the ejecta
(Jerkstrand, Fransson \& Kozma 2011). Small grains would naturally be hotter
within the ISRF, and also more prone to stochastic heating, which could imply
that not all dust mass is accounted for (the addition of a much colder
component could mimic a flatter long-wavelength tail). Small grains might
result from sputtering by the hot gas and X-rays impinging upon the ring or
even the ejecta (Larsson et al.\ 2011). While Dwek et al.\ (2010) did find an
additional population of small grains in the equatorial ring, these were much
hotter ($\sim350$ K). Hence the cold dust might be most easily explained as
due to condensation within the ejecta, but there are scenarios possible that
reduce the inferred mass to $\ll0.5$ M$_\odot$, leaving other options open.

\section{Conclusions}

We presented new images of SN\,1987A at wavelengths of 350 and 870 $\mu$m,
obtained with the APEX telescope, showing very clear detections of largely
unresolved emission. The image at 350 $\mu$m improves the constraints on the
location of the emission by an order of magnitude in area over that obtained
from {\it Herschel} observations, to well within $8^{\prime\prime}$ from the
explosion site. The far-IR and sub-mm SED is consistent with emission from
cold dust, but the temperature may be as high as $\sim33$ K if the emission
coefficient is wavelength-independent -- though this would require a
contribution from line emission to the flux within the 160- and 250-$\mu$m
{\it Herschel} bands. Smaller, non-spherical, or fluffy grains would reduce the
inferred dust mass, possibly enough to become consistent with pre-SN
production.


\acknowledgements{We kindly thank Jeonghee Rho for her constructive
referee reports. This publication is based on data acquired with the Atacama
Pathfinder Experiment (APEX). APEX is a collaboration between the Max Planck
Institut f\"ur Radioastronomie, the European Southern Observatory, and the
Onsala Space Observatory.}


\end{document}